\documentclass[aps,prc,rapid,twocolumn,superscriptaddress,nofootinbib]{revtex4-1}

\usepackage{natbib}
\usepackage{mathptmx} 
\usepackage{pstricks}
\usepackage{array}
\usepackage{tabularx}
\usepackage{graphicx}
\usepackage{color}
\usepackage{psfrag}
\usepackage{amsmath}
\usepackage{amssymb}
\usepackage{psfrag}
\usepackage{textcomp}
\usepackage{longtable}
\usepackage{multirow}
\usepackage{braket}
\usepackage[mathscr]{euscript}
\usepackage{dcolumn}
\usepackage{bm}
\usepackage{hyperref}
\hypersetup{letterpaper, colorlinks=true, urlcolor=blue, citecolor=blue, linkcolor=blue, breaklinks=true}
\usepackage{breakurl}

\newcommand{\citeref}[1]{Ref.\ \cite{#1}}

\newcommand{\meas}[2]{$#1~\textrm{#2}$}
\newcommand{\nuc}[2]{$^{#1}\textrm{#2}$}

\newcommand{\nuccharge}[3]{$^{#1}\textrm{#2}^{\left( #3 \right) }$}
\newcommand{\nucmath}[2]{^{#1}\mathrm{#2}}

\newcommand{\inchcommand}{^{\prime \prime}}

\newcommand{\rxnfull}[6]{$^{#1}\textrm{#2} \left( #3 , #4 \right) ^{#5}\textrm{#6}$}

\newcommand{\ee}[2]{#1 \times 10^{#2}}
\newcommand{\wg}{\omega \gamma}

\newcommand{\figref}[1]{Fig.\ \ref{#1}}
\newcommand{\eqnref}[1]{Eq.\ \eqref{#1}}
\newcommand{\tableref}[1]{Table~\ref{#1}}

\newcommand{\executeiffilenewer}[3]{%
\if{\Filemodnewest[1]{#1}{#2} == #2}
{\immediate\write18{#3}}\fi%
}

\newcommand{\triumf}{TRIUMF, 4004 Wesbrook Mall, Vancouver, BC V6T 2A3, Canada}
\newcommand{\york}{Department of Physics, University of York, Heslington, York YO10 5DD, UK}
\newcommand{\csm}{Department of Physics, Colorado School of Mines, 1523 Illinois Street, Golden, CO 80401, USA}

\begin{document}
\title{Strength of the $E_{\text{cm}} = 1113$ keV resonance in $^{20}${Ne}$(p, \gamma)^{21}${Na}}

\author{G. Christian}
\thanks{\href{mailto:gchristian@triumf.ca}{gchristian@triumf.ca}}
\affiliation{\triumf}

\author{D. Hutcheon}
\affiliation{\triumf}

\author{C. Akers}
\affiliation{\triumf}
\affiliation{\york}

\author{D. Connolly}
\affiliation{\csm}


\author{J. Fallis}
\affiliation{\triumf}

\author{C. Ruiz}
\affiliation{\triumf}

\date{\today}

\begin{abstract}

The $^{20}$Ne$(p, \gamma)^{21}$Na reaction is the starting point of the NeNa cycle, which is an important process for the production of intermediate mass elements. The $E_{\text{cm}} = 1113$ keV resonance plays an important role in the determination of stellar rates for this reaction since it is used to normalize experimental direct capture yields at lower energies. The commonly accepted strength of this resonance, $\omega \gamma = 1.13 \pm 0.07$ eV, has been misinterpreted as the strength in the center-of-mass frame when it is actually the strength in the laboratory frame. This has motivated a new measurement of the $E_{\text{cm}} = 1113$ keV resonance strength in $^{20}$Ne$(p, \gamma)^{21}$Na using the DRAGON recoil mass spectrometer. The DRAGON result, $0.972 \pm 0.11$ eV, is in good agreement with the accepted value when both are calculated in the same frame of reference.

\end{abstract}


\maketitle

\section{Introduction}

The \rxnfull{20}{Ne}{p}{\gamma}{21}{Na} reaction is the starting point for the NeNa cycle \cite{Rolfs1975460},
\begin{eqnarray}
\label{eq:nena_cycle}
	\nucmath{20}{Ne}(p, \gamma)
	\nucmath{21}{Na}(\beta^{+} \nu)
	\nucmath{21}{Ne}(p, \gamma) 
	\nucmath{22}{Na}(\beta^{+} \nu) & \nonumber \\*
	\nucmath{22}{Ne}(p, \gamma) 
	\nucmath{23}{Na}(p, \alpha)
	\nucmath{20}{Ne},
\end{eqnarray}
which is a key process for the nucleosynthesis of intermediate mass elements in ONe classical novae \cite{Rolfs1975460, 1999ApJ...520..347J} and the production of sodium in yellow supergiants \cite{1991ApJ...379..729P}.

The cross section of \rxnfull{20}{Ne}{p}{\gamma}{21}{Na} has been measured in the range of $E_\text{cm} = 0.35$--\meas{2.0}{MeV} ($E_{\text{lab}} = 0.37$--\meas{2.1}{MeV}) by Rolfs \textit{et al.}\ \cite{Rolfs1975460} and extrapolated to stellar energies. This experiment did not measure absolute yields and instead relied on normalization to the previously measured \meas{E_{\text{cm}} = 1113}{keV} (\meas{E_\text{lab} = 1169}{keV}) resonance as well as \rxnfull{16}{O}{p}{\gamma}{17}{F} direct capture.  The ``known'' value of the \meas{E_{\text{cm}} = 1113}{keV} resonance was taken from \citeref{Bloch1969129}, which was in turn normalized to the result of Thomas and Tanner \cite{ThomasAndTanner}. Although not explicitly stated by Thomas and Tanner, we have determined that their resonance strength measurement is presented in the laboratory frame (in which the proton is moving and the \nuc{20}{Ne} is at rest), whereas subsequent normalizations would have required the resonance strength in the center-of-mass frame.

The importance of the \meas{E_{\text{cm}} = 1113}{keV} resonance strength for direct capture normalization and the questions surrounding the Thomas and Tanner result motivate a new measurement of $\wg_{1113}.$  We have performed this measurement in inverse kinematics using the DRAGON \cite{Hutcheon2003190} recoil mass spectrometer. The new result is in good agreement with the Thomas and Tanner strength when both are calculated in the same reference frame.

\section{Thomas and Tanner Result}
\label{sec:TannerThomas}

The Thomas and Tanner measurement \cite{ThomasAndTanner} was performed by impinging a proton beam onto a target filled with natural neon gas. To measure the \nuc{21}{Na} yield, the authors used a pair of $4 \inchcommand \times 4 \inchcommand$ NaI detectors to count $\gamma$ rays resulting from positron annihilation in the target walls. The efficiency of the NaI counters was calibrated using a \nuc{22}{Na} source with activity of \meas{\sim 1}{$\mu$C}.  For \meas{1169}{keV} protons, they quote a yield of $(6.5 \pm 0.3) \times 10^{-10}$ and from this extract a resonance strength of \meas{\wg = 1.13 \pm 0.07}{eV} according to the following:

\begin{quote}
Integrating the Breit-Wigner formula for a thick target gives the yield per 
proton as $2 \pi^{2} \lambdabar^2 \omega \Gamma_{p} \Gamma_{\gamma} / \Gamma \epsilon
 \simeq 2 \pi^{2} \lambdabar^2 \omega \Gamma_{\gamma} / {\epsilon} $
 if $\Gamma_{p} \gg \Gamma_{\gamma}, $ where $\lambdabar$ is the de Broglie wavelength, $\Gamma_{p},$ $\Gamma_{\gamma}$ and $\Gamma$ the partial and total widths of the resonance, $\omega$ the statistical factor and $\epsilon$ the stopping power.
\end{quote}
We are interested in the total resonance strength, $\wg = \omega \Gamma_{p} \Gamma_{\gamma} / \Gamma,$ so we can ignore the $\Gamma_{p} \gg \Gamma_{\gamma}$ approximation and write $\wg$ where Thomas and Tanner would have written $\omega \Gamma_{\gamma}.$

Calculation details are rather sparse, and the authors do not explicitly state the reference frame in which they present $\wg$. However, close examination of the formulae makes it clear that it must be the laboratory frame. Using contemporary values for the de Broglie constants \cite{RevModPhys.27.363}, we calculate a de Broglie wavelength for \meas{1169}{keV} protons of \meas{\lambda^{2}_\text{lab} = \ee{7.0}{-24}}{cm$^{2}$.}  For the stopping power, Thomas and Tanner quote a value of \meas{\ee{(5.55 \pm 0.2)}{-15}}{eV cm$^2$} \cite{WhalingEloss}. Examination of Fig. 4 in \citeref{WhalingEloss} indicates that this is for natural neon, and presumably Thomas and Tanner would have corrected for the \nuc{20}{Ne} abundance (they explicitly state that their target used natural neon gas). Using a contemporary reference for the neon abundances \cite{PhysRev.79.450}, we calculate a \nuc{20}{Ne} effective stopping power of \meas{(6.10 \pm 0.22) \times 10^{-15}}{eV cm$^{2}$.} Taking these values and the quoted yield of $(6.5 \pm 0.3) \times 10^{-10},$ we arrive at a resonance strength of \meas{\wg = 1.13 \pm 0.07}{eV}, in exact numerical agreement with what is quoted in the paper. 

Further evidence that the resonance strength is in the laboratory frame can be found in a contemporary paper by Tanner \cite{PhysRev.114.1060}, where he states that \rxnfull{20}{Ne}{p}{\gamma}{21}{Na} direct capture yield can be normalized to the yield of the \meas{E_\text{lab} = 1169}{keV} resonance using the formula $Y_{\textrm{DC}} / Y_{\textrm{R}} = 0.29 \sigma \Delta / \wg,$ where $\sigma$ is the direct capture cross section in barns. The factor of \meas{0.29}{b$^{-1}$} is reproduced exactly by $2 / \lambda^2,$ where \meas{\lambda^2 = 7.0}{b} is the square of the laboratory de Broglie wavelength for an \meas{1169}{keV} proton \cite{RevModPhys.27.363}. This indicates that the formula for relative yields should take $\wg$ in the laboratory frame. If $\wg_{\text{cm}}$ were expected, the factor would instead be $(2 / \lambda_{\text{cm}}^{2})  [M/(m + M)],$ which evaluates to \meas{0.24}{b$^{-1}$}. Although it is not possible to reproduce Tanner's calculation of direct capture cross sections since he does not explicitly provide $Y_{\textrm{DC}} / Y_{\textrm{R}}$, the ratio of cross sections calculated with an earlier measurement of \meas{\wg_{\text{lab}} = 2}{eV} \cite{BrostromNature1947, PhysRev.71.661}
and \meas{\wg = 1.13}{eV} is equal to $1.13/2.$  Thus the two resonance strengths must both be presented in the laboratory reference frame.

To convert the resonance strength into the center-of-mass frame, we multiply by 
$\left[ M / \left( m + M \right) \right]^{3},$ where $M$ $(m)$ is the \nuc{20}{Ne} (proton) mass,
to arrive at a value of \meas{\wg = 0.975 \pm 0.060}{eV}.\footnote{
The conversion from lab to center-of-mass frame proceeds as follows:
\begin{eqnarray*}
		 \wg_{\text{lab}} & = & \frac{Y \epsilon}{2 \pi^2  \lambdabar^2_\text{lab}} 
							   =   \frac{Y \epsilon}{2 \pi^2} \frac{2 m E_{\text{lab}}}{\hbar^2} \\
		 &	 = & 
		 		\frac{Y \epsilon}{2 \pi^2} \frac{2}{\hbar^2} 	\left( \frac{M + m} {M} \mu \right) 
		 														\left( \frac{M + m}{M} E_{\text{cm}} \right) \\
		 	& = & \frac{Y \epsilon}{2 \pi^2 \lambdabar_{\text{cm}}^{2}} 
		 							\frac{M}{M + m} \left( \frac{M + m} {M} \right)^3 \\
		 	& = & \wg_\text{cm} \left( \frac{M + m} {M} \right)^3.
\end{eqnarray*}
} It may be advisable to increase this value by $3$--$4\%$ to account for more modern calculations of stopping power. For example, ATIMA \cite{AtimaOnline} and SRIM-2013 \cite{ziegler1985stopping} give stopping powers of $\ee{5.76}{-15}$ and $\ee{5.73}{-15}~${}eV~cm$^{2},$ respectively.

\section{DRAGON Measurement}
\label{sec:dragon}
The strength of the \meas{E_{\text{cm}} = 1113}{keV} resonance in \rxnfull{20}{Ne}{p}{\gamma}{21}{Na} was recently measured using the DRAGON recoil mass spectrometer \cite{Hutcheon2003190}, part of the ISAC facility at TRIUMF \cite{Laxdal2003400}. This experiment also served as commissioning of a new timestamp-based data acquisition system that will be detailed in an upcoming technical paper. An isotopically pure beam of \nuccharge{20}{Ne}{5+} was extracted from the ISAC offline microwave ion source \cite{jayamanna:02C711} and accelerated to an energy of \meas{1163}{keV/u} before being delivered to the DRAGON experimental station with an average intensity of \meas{(\ee{9.3 \pm 0.1)}{10}}{s}$^{-1}$. The \nuc{20}{Ne} beam impinged on the DRAGON windowless gas target, which was filled with H$_2$ gas at an average pressure of \meas{4.81 \pm 0.01}{Torr}. The incoming beam intensity was monitored by counting elastically scattered protons in a silicon surface barrier detector located inside the gas target at a $30^{\circ}$ angle from the beam axis. To obtain an absolute measure of the beam current, the number of recorded scatters was normalized to Faraday cup readings taken every hour.

\begin{figure}
\centering
\includegraphics{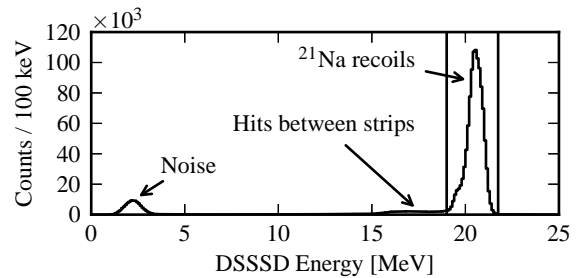}
	\caption{Sum of energy deposited in the front and back strips of the DRAGON DSSSD. The cut used in the final analysis of recoil yields is shown by the vertical lines in the figure.}
	\label{fig:recoilId}
\end{figure}

The \nuccharge{21}{Na}{9+} recoils resulting from resonant proton capture were transmitted to the end of DRAGON where they were detected in a double-sided silicon strip detector (DSSSD) \cite{Wrede2003619}. As shown in \figref{fig:recoilId}, a strong \nuc{21}{Na} peak was clearly present in the DSSSD energy sum spectrum with virtually no background from unreacted beam, allowing the resonance strength to be determined from the singles yield of \nuc{21}{Na}. As indicated in the figure, \nuc{21}{Na} events were selected by placing a cut on the main peak in the DSSSD energy spectrum.

The resonance strength was calculated using the standard formula for thick target yield in inverse kinematics \cite{rolfs1988cauldrons},
\begin{equation}
\label{eqn:wg}
	\wg = \frac{2 Y \epsilon}{\lambda_{\text{cm}}^{2}} \frac{m}{m + M},
\end{equation}
where $Y$ is the recoil yield, $\epsilon$ the lab-frame stopping power, $\lambda_{\text{cm}}$ the center-of-mass de Broglie wavelength, and $m$ $(M)$ the proton (\nuc{20}{Ne}) mass. The yield was calculated from the number of detected recoils, $n_r,$ integrated beam flux, $n_b,$ and detection efficiency, $\eta,$ as $Y = n_r / ( \eta n_b ).$ A quantitative summary of the yield calculation is given in \tableref{table:yield}.

\begin{table}
\caption{Summary of the DRAGON yield calculation. The total efficiency $\eta$ is the product of the individual efficiencies listed in the table, and the yield is calculated as $Y = n_r / ( \eta n_b ).$}
	\centering
	\begin{ruledtabular}
	\begin{tabular}{ll}
		\textit{Quantity}								&	\textit{Value}			\\ 
		\hline \\[-8pt]
		DSSSD detection efficiency						& $97.0 \pm 0.7\%$			 \cite{Wrede2003619}		 \\
		Neon 9+ charge state fraction					& $59 \pm 1\%$ 				 \cite{Engel2005491}	 	 \\
		DRAGON transmission 							& $99.9^{+0.1}_{-0.2}\%$	 \cite{EngelThesis}	 	 \\
		Microchannel plate transmission				& $76.9 \pm 0.6\%$			 \cite{Vockenhuber2009372}  \\
		Gas target transmission						& $94\%$					 					         \\
		Live time										& $95.6\%$					 				        	 \\
		Total efficiency $(\eta)$						& $39.5 \pm 0.8\%$			 					         \\
		Number of detected recoils $(n_r)$				& $\ee{(1.11 \pm 0.01)}{6}$    					         \\
		Integrated beam current $(n_b)$				& $\ee{(2.30 \pm 0.03)}{15}$	 					         \\
		Yield $(Y)$										& $\ee{(1.22 \pm 0.03)}{-9}$	 
	\end{tabular}
	\end{ruledtabular}
\label{table:yield}
\end{table}

For stopping power, $\epsilon,$ we use the published values of \citeref{Greife20041}. To account for the small difference between the present beam energy and the closest available energy in \citeref{Greife20041}, we fit the published measurements at $760$, $854$, and \meas{1156}{keV/u} with a function inspired by the Bethe-Bloch equation,
\begin{equation}
\label{eqn:stopExtrap}
	\epsilon \left( E_b \right) = a \frac{\ln \left( E_b/b \right)}{E_b},
\end{equation}
and extrapolate to \meas{E_b = 1163}{keV/u} to arrive at a final stopping power of \meas{\ee{(63.9 \pm 7.2)}{-15}}{eV cm$^{2}$}.\footnote{The best fit values of the free parameters are \meas{a = 20565.8}{eV cm$^{2}$} and \meas{b = 31.3062}{keV/u}.}  From the measured yield, stopping power and resonance parameters, we calculate a resonance strength of \meas{\wg = 0.972 \pm 0.11 }{eV}.

\begin{figure*}
    \centering
    \includegraphics{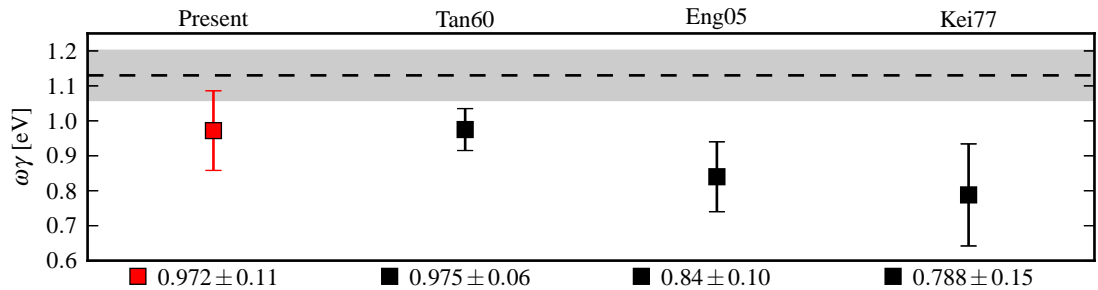}
	\caption{(Color online). Summary of strength measurements of the \meas{E_{\text{cm}} = 1113}{keV} resonance in \rxnfull{20}{Ne}{p}{\gamma}{21}{Na}. The references are Tan60 \cite{ThomasAndTanner}, Eng05 \cite{Engel2005491}, and Kei77 \cite{PhysRevC.15.579}. For comparison, the laboratory frame Thomas and Tanner result, \meas{\wg = 1.13 \pm 0.07}{eV} \cite{ThomasAndTanner}, is also displayed by the band across the top of the figure. The black dashed line denotes the central value and the grey shaded region the uncertainty.}
	\label{fig:ne20wg}
\end{figure*}

\figref{fig:ne20wg} displays the present result along with that of Thomas and Tanner \cite{ThomasAndTanner} (in the center of mass); the DRAGON commissioning experiment \cite{Engel2005491}; and the solid target measurement of Keinonen, Riihonen, and Anttila \cite{PhysRevC.15.579}. For the DRAGON commissioning experiment, we have recalculated the resonance strength using the presently employed stopping power, which has a larger error than the stopping power used originally.  This was motivated by an analysis of a large set of $^{20}\textrm{Ne} + \textrm{H}_2$ energy loss data taken subsequent to the publication of \citeref{Engel2005491}. These data indicate that the error on stopping power measurements is consistent with the value quoted by Greife \textit{et al.}\ \cite{Greife20041}, implying that the error in \citeref{Engel2005491} is underestimated.

\section{Discussion}

The Thomas and Tanner result was used by Bloch \textit{et al.}\ \cite{Bloch1969129} to normalize their measurements of individual transition strengths from excited states in \nuc{21}{Na}. For the state corresponding to the \meas{E_{\text{cm}} = 1113}{keV} resonance in $\nucmath{20}{Ne} + p$ (\meas{E^* = 3.544}{MeV,} $J^{\pi} = 5/2^{+},$ $\omega = 3$), their measured partial widths sum to \meas{\Gamma_\gamma = 0.376}{eV}, or \meas{\omega \Gamma_\gamma = 1.13}{eV}. Thus the Thomas and Tanner resonance strength must have been mistakenly taken as the center-of-mass value.  Bloch \textit{et al.}\ compare their results to a prior measurement \cite{VanderLeun1964333} that also appears to have been normalized to Thomas and Tanner without making a center-of-mass correction.

The Bloch \textit{et al.}\ results, along with an evaluation \cite{Endt19731}, were used by Rolfs \textit{et al.}\ \cite{Rolfs1975460} to normalize direct capture yields according to the formula
\begin{equation}
\label{eq:dc_res}
	\sigma_{\textrm{DC}} = \frac{1}{2} \lambda^2 \frac{m+M}{M} \frac{\wg_{1113}}{\Delta}
							 		\left( \frac{Y_{\textrm{DC}}}{Y_{\textrm{R}}} \right).
\end{equation}
From this, they calculate a cross section of \meas{\sigma = 0.63 \pm 0.08}{$\mu$b} for the \meas{\textrm{DC} \rightarrow 2425}{keV} transition at \meas{E_\text{lab} = 1.05}{MeV.}  They also perform a separate normalization to \rxnfull{16}{O}{p}{\gamma}{17}{F} direct capture, resulting in \meas{\sigma = 0.64 \pm 0.1}{$\mu$b}.  They use the weighted mean value of \meas{\sigma = 0.63 \pm 0.07}{$\mu$b} to normalize the yields of other direct capture transitions and to calculate the astrophysical $S$-factor for the reaction.

Similar normalizations of \rxnfull{20}{Ne}{p}{\gamma}{21}{Na} have been carried out in recent evaluations.  For example, the authors of \citeref{Iliadis201031} use \meas{\wg_{1113} = 1.125 \pm 0.018}{eV}, along with \rxnfull{16}{O}{p}{\gamma}{17}{F} direct capture cross sections from \citeref{Mohr201262}, to renormalize \rxnfull{20}{Ne}{p}{\gamma}{21}{Na} direct capture cross sections.  From this analysis, they recommend increasing the Rolfs \textit{et al.}\ $S$-factors by $2.9\%.$

To examine the effect of a change in $\wg_{1113}$ on the astrophysical $S$-factor, we renormalize the \meas{\textrm{DC} \rightarrow 2425}{keV} cross section using \eqnref{eq:dc_res}, taking $\wg_{1113}$ as \meas{0.931 \pm 0.045}{eV}, the weighted average of the strengths shown in \figref{fig:ne20wg}. The result is \meas{\sigma_\textrm{DC} = 0.52 \pm 0.06}{$\mu$b}, which implies a $17\%$ decrease in the $S$-factor of Rolfs \textit{et al.}  This represents the maximum potential change since inclusion of the \rxnfull{16}{O}{p}{\gamma}{17}{F} normalization would tend to increase the the \meas{\textrm{DC} \rightarrow 2425}{keV} cross section.

In conclusion, we have performed a new measurement of the strength of the \meas{E_\text{cm} = 1113}{keV} resonance in \rxnfull{20}{Ne}{p}{\gamma}{21}{Na} using the DRAGON recoil mass spectrometer.  This was motivated by the discovery that the prior accepted value, \meas{\wg = 1.13 \pm 0.07}{eV} \cite{ThomasAndTanner}, is the strength in the laboratory frame instead of the center-of-mass frame.  Subsequent experiments \cite{Rolfs1975460, Bloch1969129, VanderLeun1964333} have used the result of \citeref{ThomasAndTanner} for absolute normalization without making the necessary conversion into the center of mass. This has resulted in an overestimation of the \rxnfull{20}{Ne}{p}{\gamma}{21}{Na} $S$-factor by as much as $17 \%.$ In the future, it would be beneficial to perform stellar model calculations to investigate the impact of such a change on isotopic abundances.

\begin{acknowledgments}
The authors are grateful to the ISAC staff at TRIUMF for delivery of a high quality \nuc{20}{Ne} beam during the DRAGON experiment, as well as members of the TRIUMF data-acquisision group, K. Olchanski and P. Amaudruz, for their efforts in developing the new DRAGON data-acquisition system.  We also thank C. Illiadis for a careful reading of the manuscript and for helpful comments and suggestions.  This work was supported by the Natural Sciences and Engineering Research Council of Canada.
\end{acknowledgments}


\begin{thebibliography}{27}%
\makeatletter
\providecommand \@ifxundefined [1]{%
 \@ifx{#1\undefined}
}%
\providecommand \@ifnum [1]{%
 \ifnum #1\expandafter \@firstoftwo
 \else \expandafter \@secondoftwo
 \fi
}%
\providecommand \@ifx [1]{%
 \ifx #1\expandafter \@firstoftwo
 \else \expandafter \@secondoftwo
 \fi
}%
\providecommand \natexlab [1]{#1}%
\providecommand \enquote  [1]{``#1''}%
\providecommand \bibnamefont  [1]{#1}%
\providecommand \bibfnamefont [1]{#1}%
\providecommand \citenamefont [1]{#1}%
\providecommand \href@noop [0]{\@secondoftwo}%
\providecommand \href [0]{\begingroup \@sanitize@url \@href}%
\providecommand \@href[1]{\@@startlink{#1}\@@href}%
\providecommand \@@href[1]{\endgroup#1\@@endlink}%
\providecommand \@sanitize@url [0]{\catcode `\\12\catcode `\$12\catcode
  `\&12\catcode `\#12\catcode `\^12\catcode `\_12\catcode `\%12\relax}%
\providecommand \@@startlink[1]{}%
\providecommand \@@endlink[0]{}%
\providecommand \url  [0]{\begingroup\@sanitize@url \@url }%
\providecommand \@url [1]{\endgroup\@href {#1}{\urlprefix }}%
\providecommand \urlprefix  [0]{URL }%
\providecommand \Eprint [0]{\href }%
\providecommand \doibase [0]{http://dx.doi.org/}%
\providecommand \selectlanguage [0]{\@gobble}%
\providecommand \bibinfo  [0]{\@secondoftwo}%
\providecommand \bibfield  [0]{\@secondoftwo}%
\providecommand \translation [1]{[#1]}%
\providecommand \BibitemOpen [0]{}%
\providecommand \bibitemStop [0]{}%
\providecommand \bibitemNoStop [0]{.\EOS\space}%
\providecommand \EOS [0]{\spacefactor3000\relax}%
\providecommand \BibitemShut  [1]{\csname bibitem#1\endcsname}%
\let\auto@bib@innerbib\@empty
\bibitem [{\citenamefont {Rolfs}\ \emph {et~al.}(1975)\citenamefont {Rolfs},
  \citenamefont {Rodney}, \citenamefont {Shapiro},\ and\ \citenamefont
  {Winkler}}]{Rolfs1975460}%
  \BibitemOpen
  \bibfield  {author} {\bibinfo {author} {\bibfnamefont {C.}~\bibnamefont
  {Rolfs}}, \bibinfo {author} {\bibfnamefont {W.}~\bibnamefont {Rodney}},
  \bibinfo {author} {\bibfnamefont {M.}~\bibnamefont {Shapiro}}, \ and\
  \bibinfo {author} {\bibfnamefont {H.}~\bibnamefont {Winkler}},\ }\href
  {\doibase http://dx.doi.org/10.1016/0375-9474(75)90398-X} {\bibfield
  {journal} {\bibinfo  {journal} {Nucl. Phys. A}\ }\textbf {\bibinfo {volume}
  {241}},\ \bibinfo {pages} {460 } (\bibinfo {year} {1975})}\BibitemShut
  {NoStop}%
\bibitem [{\citenamefont {{Jos{\'e}}}\ \emph {et~al.}(1999)\citenamefont
  {{Jos{\'e}}}, \citenamefont {{Coc}},\ and\ \citenamefont
  {{Hernanz}}}]{1999ApJ...520..347J}%
  \BibitemOpen
  \bibfield  {author} {\bibinfo {author} {\bibfnamefont {J.}~\bibnamefont
  {{Jos{\'e}}}}, \bibinfo {author} {\bibfnamefont {A.}~\bibnamefont {{Coc}}}, \
  and\ \bibinfo {author} {\bibfnamefont {M.}~\bibnamefont {{Hernanz}}},\ }\href
  {\doibase 10.1086/307445} {\bibfield  {journal} {\bibinfo  {journal}
  {Astrophys. J}\ }\textbf {\bibinfo {volume} {520}},\ \bibinfo {pages} {347}
  (\bibinfo {year} {1999})}\BibitemShut {NoStop}%
\bibitem [{\citenamefont {{Prantzos}}\ \emph {et~al.}(1991)\citenamefont
  {{Prantzos}}, \citenamefont {{Coc}},\ and\ \citenamefont
  {{Thibaud}}}]{1991ApJ...379..729P}%
  \BibitemOpen
  \bibfield  {author} {\bibinfo {author} {\bibfnamefont {N.}~\bibnamefont
  {{Prantzos}}}, \bibinfo {author} {\bibfnamefont {A.}~\bibnamefont {{Coc}}}, \
  and\ \bibinfo {author} {\bibfnamefont {J.~P.}\ \bibnamefont {{Thibaud}}},\
  }\href {\doibase 10.1086/170548} {\bibfield  {journal} {\bibinfo  {journal}
  {Astrophys. J}\ }\textbf {\bibinfo {volume} {379}},\ \bibinfo {pages} {729}
  (\bibinfo {year} {1991})}\BibitemShut {NoStop}%
\bibitem [{\citenamefont {Bloch}\ \emph {et~al.}(1969)\citenamefont {Bloch},
  \citenamefont {Knellwolf},\ and\ \citenamefont {Pixley}}]{Bloch1969129}%
  \BibitemOpen
  \bibfield  {author} {\bibinfo {author} {\bibfnamefont {R.}~\bibnamefont
  {Bloch}}, \bibinfo {author} {\bibfnamefont {T.}~\bibnamefont {Knellwolf}}, \
  and\ \bibinfo {author} {\bibfnamefont {R.}~\bibnamefont {Pixley}},\ }\href
  {\doibase http://dx.doi.org/10.1016/0375-9474(69)90894-X} {\bibfield
  {journal} {\bibinfo  {journal} {Nucl. Phys. A}\ }\textbf {\bibinfo {volume}
  {123}},\ \bibinfo {pages} {129 } (\bibinfo {year} {1969})}\BibitemShut
  {NoStop}%
\bibitem [{\citenamefont {Thomas}\ and\ \citenamefont
  {Tanner}(1960)}]{ThomasAndTanner}%
  \BibitemOpen
  \bibfield  {author} {\bibinfo {author} {\bibfnamefont {G.~C.}\ \bibnamefont
  {Thomas}}\ and\ \bibinfo {author} {\bibfnamefont {N.~W.}\ \bibnamefont
  {Tanner}},\ }\href {\doibase 10.1088/0370-1328/75/4/303} {\bibfield
  {journal} {\bibinfo  {journal} {Proc. Phys. Soc.}\ }\textbf {\bibinfo
  {volume} {75}},\ \bibinfo {pages} {498 } (\bibinfo {year}
  {1960})}\BibitemShut {NoStop}%
\bibitem [{\citenamefont {Hutcheon}\ \emph {et~al.}(2003)\citenamefont
  {Hutcheon} \emph {et~al.}}]{Hutcheon2003190}%
  \BibitemOpen
  \bibfield  {author} {\bibinfo {author} {\bibfnamefont {D.}~\bibnamefont
  {Hutcheon}} \emph {et~al.},\ }\href {\doibase 10.1016/S0168-9002(02)01990-3}
  {\bibfield  {journal} {\bibinfo  {journal} {Nucl. Instr. Meth. in Phys. Res.
  A}\ }\textbf {\bibinfo {volume} {498}},\ \bibinfo {pages} {190 } (\bibinfo
  {year} {2003})}\BibitemShut {NoStop}%
\bibitem [{\citenamefont {Cohen}\ \emph {et~al.}(1955)\citenamefont {Cohen},
  \citenamefont {DuMond}, \citenamefont {Layton},\ and\ \citenamefont
  {Rollett}}]{RevModPhys.27.363}%
  \BibitemOpen
  \bibfield  {author} {\bibinfo {author} {\bibfnamefont {E.~R.}\ \bibnamefont
  {Cohen}}, \bibinfo {author} {\bibfnamefont {J.~W.}\ \bibnamefont {DuMond}},
  \bibinfo {author} {\bibfnamefont {T.~W.}\ \bibnamefont {Layton}}, \ and\
  \bibinfo {author} {\bibfnamefont {J.~S.}\ \bibnamefont {Rollett}},\ }\href
  {\doibase 10.1103/RevModPhys.27.363} {\bibfield  {journal} {\bibinfo
  {journal} {Rev. Mod. Phys.}\ }\textbf {\bibinfo {volume} {27}},\ \bibinfo
  {pages} {363} (\bibinfo {year} {1955})},\ \bibinfo {note} {the de {B}roglie
  wavelength for protons (nonrelativistic approximation) is quoted as
  $\lambda_{p} = (2.86202 \pm 0.00004) \times 10^{-9}~ \textrm{cm eV}^{1/2}~
  \textrm{E}^{-1/2}$}\BibitemShut {NoStop}%
\bibitem [{\citenamefont {Whaling}(1958)}]{WhalingEloss}%
  \BibitemOpen
  \bibfield  {author} {\bibinfo {author} {\bibfnamefont {W.}~\bibnamefont
  {Whaling}},\ }in\ \href {\doibase 10.1007/978-3-642-45898-9_5} {\emph
  {\bibinfo {booktitle} {Corpuscles and Radiation in Matter II / Korpuskeln und
  Strahlung in Materie II}}},\ \bibinfo {series} {Encyclopedia of Physics /
  Handbuch der Physik}, Vol.\ \bibinfo {volume} {6 / 34},\ \bibinfo {editor}
  {edited by\ \bibinfo {editor} {\bibfnamefont {S.}~\bibnamefont
  {Fl{\"u}gge}}}\ (\bibinfo  {publisher} {Springer Berlin Heidelberg},\
  \bibinfo {year} {1958})\ pp.\ \bibinfo {pages} {193--217}\BibitemShut
  {NoStop}%
\bibitem [{\citenamefont {Nier}(1950)}]{PhysRev.79.450}%
  \BibitemOpen
  \bibfield  {author} {\bibinfo {author} {\bibfnamefont {A.~O.}\ \bibnamefont
  {Nier}},\ }\href {\doibase 10.1103/PhysRev.79.450} {\bibfield  {journal}
  {\bibinfo  {journal} {Phys. Rev.}\ }\textbf {\bibinfo {volume} {79}},\
  \bibinfo {pages} {450} (\bibinfo {year} {1950})},\ \bibinfo {note} {the
  $^{20}$Ne abundance is quoted as $90.94 \pm 0.04\%$}\BibitemShut {NoStop}%
\bibitem [{\citenamefont {Tanner}(1959)}]{PhysRev.114.1060}%
  \BibitemOpen
  \bibfield  {author} {\bibinfo {author} {\bibfnamefont {N.}~\bibnamefont
  {Tanner}},\ }\href {\doibase 10.1103/PhysRev.114.1060} {\bibfield  {journal}
  {\bibinfo  {journal} {Phys. Rev.}\ }\textbf {\bibinfo {volume} {114}},\
  \bibinfo {pages} {1060} (\bibinfo {year} {1959})}\BibitemShut {NoStop}%
\bibitem [{\citenamefont {Brostr\"om}\ \emph
  {et~al.}(1947{\natexlab{a}})\citenamefont {Brostr\"om}, \citenamefont
  {Huus},\ and\ \citenamefont {Koch}}]{BrostromNature1947}%
  \BibitemOpen
  \bibfield  {author} {\bibinfo {author} {\bibfnamefont {K.~J.}\ \bibnamefont
  {Brostr\"om}}, \bibinfo {author} {\bibfnamefont {T.}~\bibnamefont {Huus}}, \
  and\ \bibinfo {author} {\bibfnamefont {J.}~\bibnamefont {Koch}},\ }\href
  {\doibase 10.1038/160498b0} {\bibfield  {journal} {\bibinfo  {journal}
  {Nature}\ }\textbf {\bibinfo {volume} {160}},\ \bibinfo {pages} {498}
  (\bibinfo {year} {1947}{\natexlab{a}})}\BibitemShut {NoStop}%
\bibitem [{\citenamefont {Brostr\"om}\ \emph
  {et~al.}(1947{\natexlab{b}})\citenamefont {Brostr\"om}, \citenamefont
  {Huus},\ and\ \citenamefont {Tangen}}]{PhysRev.71.661}%
  \BibitemOpen
  \bibfield  {author} {\bibinfo {author} {\bibfnamefont {K.~J.}\ \bibnamefont
  {Brostr\"om}}, \bibinfo {author} {\bibfnamefont {T.}~\bibnamefont {Huus}}, \
  and\ \bibinfo {author} {\bibfnamefont {R.}~\bibnamefont {Tangen}},\ }\href
  {\doibase 10.1103/PhysRev.71.661} {\bibfield  {journal} {\bibinfo  {journal}
  {Phys. Rev.}\ }\textbf {\bibinfo {volume} {71}},\ \bibinfo {pages} {661}
  (\bibinfo {year} {1947}{\natexlab{b}})},\ \bibinfo {note} {this paper
  presents a formula for $\omega \gamma$ that is explicitly in the lab frame
  and is cited by Ref.\ \cite{BrostromNature1947} as the source of resonance
  strength calculation from yield measurements.}\BibitemShut {Stop}%
\bibitem [{\citenamefont {Weick}(2011)}]{AtimaOnline}%
  \BibitemOpen
  \bibfield  {author} {\bibinfo {author} {\bibfnamefont {H.}~\bibnamefont
  {Weick}},\ }\href@noop {} {\enquote {\bibinfo {title} {{ATIMA}},}\ }
  (\bibinfo {year} {2011}),\ \bibinfo {note}
  {\url{http://web-docs.gsi.de/~weick/atima/}}\BibitemShut {NoStop}%
\bibitem [{\citenamefont {Ziegler}(1985)}]{ziegler1985stopping}%
  \BibitemOpen
  \bibfield  {author} {\bibinfo {author} {\bibfnamefont {J.~F.}\ \bibnamefont
  {Ziegler}},\ }\href@noop {} {\emph {\bibinfo {title} {The Stopping and Range
  of Ions in Solids}}},\ Stopping and Range of Ions in Matter, Vol 1\ (\bibinfo
   {publisher} {Pergamon Press},\ \bibinfo {year} {1985})\ \bibinfo {note}
  {\url{http://srim.org}}\BibitemShut {NoStop}%
\bibitem [{\citenamefont {Laxdal}(2003)}]{Laxdal2003400}%
  \BibitemOpen
  \bibfield  {author} {\bibinfo {author} {\bibfnamefont {R.}~\bibnamefont
  {Laxdal}},\ }\href {\doibase 10.1016/S0168-583X(02)02104-3} {\bibfield
  {journal} {\bibinfo  {journal} {Nucl. Instr. Meth. in Phys. Res. B}\ }\textbf
  {\bibinfo {volume} {204}},\ \bibinfo {pages} {400 } (\bibinfo {year}
  {2003})}\BibitemShut {NoStop}%
\bibitem [{\citenamefont {Jayamanna}\ \emph {et~al.}(2008)\citenamefont
  {Jayamanna}, \citenamefont {Ames}, \citenamefont {Cojocaru}, \citenamefont
  {Baartman}, \citenamefont {Bricault}, \citenamefont {Dube}, \citenamefont
  {Laxdal}, \citenamefont {Marchetto}, \citenamefont {MacDonald}, \citenamefont
  {Schmor}, \citenamefont {Wight},\ and\ \citenamefont
  {Yuan}}]{jayamanna:02C711}%
  \BibitemOpen
  \bibfield  {author} {\bibinfo {author} {\bibfnamefont {K.}~\bibnamefont
  {Jayamanna}}, \bibinfo {author} {\bibfnamefont {F.}~\bibnamefont {Ames}},
  \bibinfo {author} {\bibfnamefont {G.}~\bibnamefont {Cojocaru}}, \bibinfo
  {author} {\bibfnamefont {R.}~\bibnamefont {Baartman}}, \bibinfo {author}
  {\bibfnamefont {P.}~\bibnamefont {Bricault}}, \bibinfo {author}
  {\bibfnamefont {R.}~\bibnamefont {Dube}}, \bibinfo {author} {\bibfnamefont
  {R.}~\bibnamefont {Laxdal}}, \bibinfo {author} {\bibfnamefont
  {M.}~\bibnamefont {Marchetto}}, \bibinfo {author} {\bibfnamefont
  {M.}~\bibnamefont {MacDonald}}, \bibinfo {author} {\bibfnamefont
  {P.}~\bibnamefont {Schmor}}, \bibinfo {author} {\bibfnamefont
  {G.}~\bibnamefont {Wight}}, \ and\ \bibinfo {author} {\bibfnamefont
  {D.}~\bibnamefont {Yuan}},\ }\href {\doibase 10.1063/1.2816928} {\bibfield
  {journal} {\bibinfo  {journal} {Rev. Sci. Instr.}\ }\textbf {\bibinfo
  {volume} {79}},\ \bibinfo {eid} {02C711} (\bibinfo {year}
  {2008})}\BibitemShut {NoStop}%
\bibitem [{\citenamefont {Wrede}\ \emph {et~al.}(2003)\citenamefont {Wrede},
  \citenamefont {Hussein}, \citenamefont {Rogers},\ and\ \citenamefont
  {D’Auria}}]{Wrede2003619}%
  \BibitemOpen
  \bibfield  {author} {\bibinfo {author} {\bibfnamefont {C.}~\bibnamefont
  {Wrede}}, \bibinfo {author} {\bibfnamefont {A.}~\bibnamefont {Hussein}},
  \bibinfo {author} {\bibfnamefont {J.~G.}\ \bibnamefont {Rogers}}, \ and\
  \bibinfo {author} {\bibfnamefont {J.}~\bibnamefont {D’Auria}},\ }\href
  {\doibase 10.1016/S0168-583X(02)02140-7} {\bibfield  {journal} {\bibinfo
  {journal} {Nucl. Instr. Meth. in Phys. Res. B}\ }\textbf {\bibinfo {volume}
  {204}},\ \bibinfo {pages} {619 } (\bibinfo {year} {2003})}\BibitemShut
  {NoStop}%
\bibitem [{\citenamefont {Rolfs}\ and\ \citenamefont
  {Rodney}(1988)}]{rolfs1988cauldrons}%
  \BibitemOpen
  \bibfield  {author} {\bibinfo {author} {\bibfnamefont {C.}~\bibnamefont
  {Rolfs}}\ and\ \bibinfo {author} {\bibfnamefont {W.}~\bibnamefont {Rodney}},\
  }\href@noop {} {\emph {\bibinfo {title} {Cauldrons in the Cosmos: Nuclear
  Astrophysics}}},\ Theoretical Astrophysics\ (\bibinfo  {publisher}
  {University of Chicago Press},\ \bibinfo {year} {1988})\BibitemShut {NoStop}%
\bibitem [{\citenamefont {Engel}\ \emph {et~al.}(2005)\citenamefont {Engel}
  \emph {et~al.}}]{Engel2005491}%
  \BibitemOpen
  \bibfield  {author} {\bibinfo {author} {\bibfnamefont {S.}~\bibnamefont
  {Engel}} \emph {et~al.},\ }\href {\doibase 10.1016/j.nima.2005.07.029}
  {\bibfield  {journal} {\bibinfo  {journal} {Nucl. Instr. Meth. in Phys. Res.
  A}\ }\textbf {\bibinfo {volume} {553}},\ \bibinfo {pages} {491 } (\bibinfo
  {year} {2005})}\BibitemShut {NoStop}%
\bibitem [{\citenamefont {Engel}(2003)}]{EngelThesis}%
  \BibitemOpen
  \bibfield  {author} {\bibinfo {author} {\bibfnamefont {S.}~\bibnamefont
  {Engel}},\ }\emph {\bibinfo {title} {Awakening of the {DRAGON}:
  {C}ommissioning of the {DRAGON} recoil separator facility and first studies
  on the $^{21}${Na}$(\alpha, p)^{22}${Mg} reaction}},\ \href
  {http://astro.triumf.ca/sites/default/files/sabine_thesis.pdf} {Ph.D.
  thesis},\ \bibinfo  {school} {Ruhr-Universit{\"a}t Bochum}, \bibinfo
  {address} {Bochum, Germany} (\bibinfo {year} {2003})\BibitemShut {NoStop}%
\bibitem [{\citenamefont {Vockenhuber}\ \emph {et~al.}(2009)\citenamefont
  {Vockenhuber}, \citenamefont {Erikson}, \citenamefont {Buchmann},
  \citenamefont {Greife}, \citenamefont {Hager}, \citenamefont {Hutcheon},
  \citenamefont {Lamey}, \citenamefont {Machule}, \citenamefont {Ottewell},
  \citenamefont {Ruiz},\ and\ \citenamefont {Ruprecht}}]{Vockenhuber2009372}%
  \BibitemOpen
  \bibfield  {author} {\bibinfo {author} {\bibfnamefont {C.}~\bibnamefont
  {Vockenhuber}}, \bibinfo {author} {\bibfnamefont {L.}~\bibnamefont
  {Erikson}}, \bibinfo {author} {\bibfnamefont {L.}~\bibnamefont {Buchmann}},
  \bibinfo {author} {\bibfnamefont {U.}~\bibnamefont {Greife}}, \bibinfo
  {author} {\bibfnamefont {U.}~\bibnamefont {Hager}}, \bibinfo {author}
  {\bibfnamefont {D.}~\bibnamefont {Hutcheon}}, \bibinfo {author}
  {\bibfnamefont {M.}~\bibnamefont {Lamey}}, \bibinfo {author} {\bibfnamefont
  {P.}~\bibnamefont {Machule}}, \bibinfo {author} {\bibfnamefont
  {D.}~\bibnamefont {Ottewell}}, \bibinfo {author} {\bibfnamefont
  {C.}~\bibnamefont {Ruiz}}, \ and\ \bibinfo {author} {\bibfnamefont
  {G.}~\bibnamefont {Ruprecht}},\ }\href {\doibase 10.1016/j.nima.2009.02.016}
  {\bibfield  {journal} {\bibinfo  {journal} {Nucl. Instr. Meth. in Phys. Res.
  A}\ }\textbf {\bibinfo {volume} {603}},\ \bibinfo {pages} {372 } (\bibinfo
  {year} {2009})}\BibitemShut {NoStop}%
\bibitem [{\citenamefont {Greife}\ \emph {et~al.}(2004)\citenamefont {Greife}
  \emph {et~al.}}]{Greife20041}%
  \BibitemOpen
  \bibfield  {author} {\bibinfo {author} {\bibfnamefont {U.}~\bibnamefont
  {Greife}} \emph {et~al.},\ }\href {\doibase 10.1016/j.nimb.2003.09.042}
  {\bibfield  {journal} {\bibinfo  {journal} {Nucl. Instr. Meth. in Phys. Res.
  B}\ }\textbf {\bibinfo {volume} {217}},\ \bibinfo {pages} {1 } (\bibinfo
  {year} {2004})}\BibitemShut {NoStop}%
\bibitem [{\citenamefont {Keinonen}\ \emph {et~al.}(1977)\citenamefont
  {Keinonen}, \citenamefont {Riihonen},\ and\ \citenamefont
  {Anttila}}]{PhysRevC.15.579}%
  \BibitemOpen
  \bibfield  {author} {\bibinfo {author} {\bibfnamefont {J.}~\bibnamefont
  {Keinonen}}, \bibinfo {author} {\bibfnamefont {M.}~\bibnamefont {Riihonen}},
  \ and\ \bibinfo {author} {\bibfnamefont {A.}~\bibnamefont {Anttila}},\ }\href
  {\doibase 10.1103/PhysRevC.15.579} {\bibfield  {journal} {\bibinfo  {journal}
  {Phys. Rev. C}\ }\textbf {\bibinfo {volume} {15}},\ \bibinfo {pages} {579}
  (\bibinfo {year} {1977})}\BibitemShut {NoStop}%
\bibitem [{\citenamefont {der Leun}\ and\ \citenamefont
  {Mouton}(1964)}]{VanderLeun1964333}%
  \BibitemOpen
  \bibfield  {author} {\bibinfo {author} {\bibfnamefont {C.~V.}\ \bibnamefont
  {der Leun}}\ and\ \bibinfo {author} {\bibfnamefont {W.~L.}\ \bibnamefont
  {Mouton}},\ }\href {\doibase 10.1016/0031-8914(64)90006-0} {\bibfield
  {journal} {\bibinfo  {journal} {Physica}\ }\textbf {\bibinfo {volume} {30}},\
  \bibinfo {pages} {333 } (\bibinfo {year} {1964})}\BibitemShut {NoStop}%
\bibitem [{\citenamefont {Endt}\ and\ \citenamefont {van~dèr
  Leun}(1973)}]{Endt19731}%
  \BibitemOpen
  \bibfield  {author} {\bibinfo {author} {\bibfnamefont {P.}~\bibnamefont
  {Endt}}\ and\ \bibinfo {author} {\bibfnamefont {C.}~\bibnamefont {van~dèr
  Leun}},\ }\href {\doibase http://dx.doi.org/10.1016/0375-9474(73)91131-7}
  {\bibfield  {journal} {\bibinfo  {journal} {Nucl. Phys. A}\ }\textbf
  {\bibinfo {volume} {214}},\ \bibinfo {pages} {1 } (\bibinfo {year}
  {1973})}\BibitemShut {NoStop}%
\bibitem [{\citenamefont {Iliadis}\ \emph {et~al.}(2010)\citenamefont
  {Iliadis}, \citenamefont {Longland}, \citenamefont {Champagne}, \citenamefont
  {Coc},\ and\ \citenamefont {Fitzgerald}}]{Iliadis201031}%
  \BibitemOpen
  \bibfield  {author} {\bibinfo {author} {\bibfnamefont {C.}~\bibnamefont
  {Iliadis}}, \bibinfo {author} {\bibfnamefont {R.}~\bibnamefont {Longland}},
  \bibinfo {author} {\bibfnamefont {A.}~\bibnamefont {Champagne}}, \bibinfo
  {author} {\bibfnamefont {A.}~\bibnamefont {Coc}}, \ and\ \bibinfo {author}
  {\bibfnamefont {R.}~\bibnamefont {Fitzgerald}},\ }\href {\doibase
  http://dx.doi.org/10.1016/j.nuclphysa.2010.04.009} {\bibfield  {journal}
  {\bibinfo  {journal} {Nucl. Phys. A}\ }\textbf {\bibinfo {volume} {841}},\
  \bibinfo {pages} {31 } (\bibinfo {year} {2010})}\BibitemShut {NoStop}%
\bibitem [{\citenamefont {Mohr}\ and\ \citenamefont
  {Iliadis}(2012)}]{Mohr201262}%
  \BibitemOpen
  \bibfield  {author} {\bibinfo {author} {\bibfnamefont {P.}~\bibnamefont
  {Mohr}}\ and\ \bibinfo {author} {\bibfnamefont {C.}~\bibnamefont {Iliadis}},\
  }\href {\doibase 10.1016/j.nima.2012.05.084} {\bibfield  {journal} {\bibinfo
  {journal} {Nucl. Instrum. Meth. in Phys. Res. A}\ }\textbf {\bibinfo {volume}
  {688}},\ \bibinfo {pages} {62 } (\bibinfo {year} {2012})}\BibitemShut
  {NoStop}%
\end{thebibliography}

%

 \end{document}